\documentclass[aps,reprint,amsmath,amssymb,prapplied]{revtex4-2}

\usepackage{graphicx}% Include figure files
\usepackage{epstopdf}
\usepackage{dcolumn}% Align table columns on decimal point
\usepackage{bm}% bold math
\usepackage{layouts}
\usepackage{ulem}
\usepackage{verbatim}

\usepackage{amsmath}
\usepackage{hyperref}% add hypertext capabilities
\hypersetup{
	colorlinks   = true, %Colours links instead of ugly boxes
	urlcolor     = blue, %Colour for external hyperlinks
	linkcolor    = blue, %Colour of internal links
	citecolor   = blue %Colour of citations, could be ``red''
}
\usepackage[dvipsnames]{xcolor}
\usepackage{float}

\usepackage{siunitx}

\begin{document}

\title{In-Situ Differential-Light-Shift Cancellation for Trapped-Atom Clocks}

\newcommand{\red}[1]{\textcolor{red}{#1}} %Alex
\newcommand{\green}[1]{\textcolor{green}{#1}} %Simon
\newcommand{\blue}[1]{\textcolor{blue}{#1}} %Carsten
\newcommand{\del}[1]{\textcolor{blue}{\sout{#1}}}
\newcommand{\rep}[2]{\textcolor{blue}{\sout{#1} #2}}

\author{Jan Simon Haase$^{1,2}$} 
\email[Author to whom correspondence should be addressed. ]{\mbox{haase@iqo.uni-hannover.de}}
\author{Alexander Fieguth$^2$}
\email[Author to whom correspondence should be addressed. ]{\mbox{alexander.fieguth@dlr.de}}

\author{Igor Bröckel$^2$}
\author{Jens Kruse$^2$}
\author{Carsten Klempt$^{1,2}$}

\affiliation{$^1$Institut für Quantenoptik, Leibniz Universit\"at Hannover, Welfengarten 1, D-30167 Hannover, Germany  \\ $^2$Deutsches Zentrum f\"ur Luft- und Raumfahrt e.V. (DLR), Institut f\"ur Satellitengeod\"asie und Inertialsensorik (DLR-SI), Callinstraße 30b, D-30167 Hannover, Germany}

\date{\today}

\begin{abstract}
Differential light shifts (DLS) induced by optical trapping fields fundamentally limit the stability and accuracy of trapped-atom microwave clocks. We demonstrate an in-situ method to cancel DLS by simultaneously interrogating multiple spatially separated atomic ensembles at different trap intensities generated from a common light source. By operating the ensembles at set intensity ratios and performing Ramsey spectroscopy, the intensity-dependent frequency shifts are measured within each experimental cycle and extrapolated to the zero-intensity limit. This approach effectively enables shot-to-shot determination of a DLS-free frequency without requiring magic wavelengths or species-specific cancellation schemes. We validate the method for Rb atoms trapped in time-averaged potentials by introducing controlled variations of the total trap power and show that the extrapolated frequency remains insensitive to these fluctuations. The technique is general and can be extended to other systematic shifts, providing a scalable route toward improved stability and accuracy in compact trapped-atom clocks and related quantum sensors relying on optical dipole traps.
\end{abstract}

\maketitle

\section{Introduction}
Atomic ensembles serve as probes for a wide range of sensing applications, including the interferometric measurement of inertia, external fields, and frequency or time.
The resolution of atom‑interferometric measurements improves linearly with the interrogation time.
When the atomic ensembles are in free fall, the pursuit of higher sensitivities leads to increasingly large instruments, such as long‑baseline atom interferometers~\cite{Abdalla2025} or microwave fountain clocks~\cite{Wynands2005}.
In principle, Ramsey interferometry on ultracold atomic ensembles can be performed also with trapped atoms, where the interrogation time is no longer bounded by the free-fall distance and thus by the size of the instrument.
However, the atoms' interaction with the trapping fields typically deteriorates the measurement.
If laser light is used for trapping and the two involved atomic levels exhibit a difference in polarizability, the trapping light itself induces a differential light shift (DLS), which results in a corresponding shift of the reference transition.
For the realization of a microwave frequency standard, where a hyperfine transition is probed via microwaves, the DLS produces three challenges:
(i) The necessarily inhomogeneous trapping field leads to an inhomogeneous frequency shift.
The atomic ensemble in the trapping field thus averages over different shifts, which leads to decoherence and a reduction of the interrogation time.
(ii) Fluctuations of the trapping field intensity result in fluctuations of the probed reference frequency, which limits the frequency stability.
(iii) Uncontrolled fluctuations and drifts of the trapping field intensity prohibit referencing to the unperturbed reference frequency and limit the accuracy. Notably, effects (i) and (ii) limit the interrogation time in distinct ways. Decoherence (i) bounds the interrogation time that can be realized at all, while the intensity-noise floor (ii) bounds the interrogation time that is worth realizing: once the single-shot resolution is limited by DLS fluctuations rather than by projection noise, a longer Ramsey time no longer improves the frequency resolution. Extending the usable interrogation time therefore requires both a sufficient coherence time and the suppression of intensity-induced frequency noise.

For optical lattice clocks all these effects can be avoided by choosing a magic wavelength for the trapping field, where the DLS vanishes in first order~\cite{Katori03}.
The effect however reappears on a lower level due to higher-order contributions to the DLS~\cite{Bothwell2025, Aeppli2024, mcgrew2018}.
For microwave clocks, such magic wavelengths typically do not exist \cite{Rosenbusch2009}, therefore the DLS effectively limits the development of compact, trapped-atom microwave references regarding stability and accuracy. 

For microwave frequency standards in magnetic traps, extended interrogation times have been demonstrated through a partial cancellation of magnetic and collisional shifts~\cite{Szmuk2015} and by exploiting spin‑self‑rephasing~\cite{Deutsch2010}.
In optical traps, spin‑self‑rephasing can likewise be exploited to suppress decoherence effects~\cite{KleineBuening2011}. For some specific atomic systems, other cancellation strategies using polarization tuning and magic configurations have been demonstrated \cite{Chicireanu2011}. 
However, in optically trapped systems these strategies only resolve the decoherence limitation (i), but are still susceptible to stability (ii) and accuracy (iii). 

The magnitude of these effects can be estimated from the measurement of the DLS per intensity for the Rb clock transition at 1064 nm, $\delta \alpha / 2 \pi = -2.2(3)$\,Hz/(kW\,cm$^{-2}$) \cite{KleineBuening2011}. 
At the per-trap intensities used in Ref. \cite{KleineBuening2011}, this corresponds to a shift of several tens of Hz, so that a relative intensity stability of order $10^{-4}$ still leaves a DLS-limited fractional instability in the $10^{-13}$ range, and slow intensity drifts bias the frequency without averaging down. 
Beyond the trap intensity itself, the light shift experienced by the atoms also varies through beam pointing (something an intensity stabilization cannot account for, although it changes the intensity at the trap location) and ensemble temperature. These contributions are comparable in size and considerably harder to characterize \cite{KleineBuening2011}.

In this article, we propose and experimentally demonstrate a concept to avoid the DLS effects (ii) and (iii).
Multiple atomic ensembles are trapped using the same light source, but set to different intensities at each ensemble location.
Thereby, we can quantify the DLS and infer the unperturbed reference frequency for each individual spectroscopy run. Since the extrapolation removes the linear, common-mode contributions of all sources discussed above, neither knowledge of $\delta\alpha$ nor of the absolute intensity at the atoms is required, and the scheme can be applied at any trap depth.
Our realization is based on microwave spectroscopy of ultracold rubidium atoms in three optical traps \cite{Anton24,Haase2025} that are formed by time-averaged potentials \cite{Roy16}.
Those potentials also enable a suppression of the decoherence effect (i) by a creation of shallow box-like trapping potentials.
We outline how our proof of concept can be exploited for the construction of compact atomic clocks with high stability and accuracy.

\begin{figure*}[htbp]

    \centering
    \includegraphics[width=0.8\linewidth]{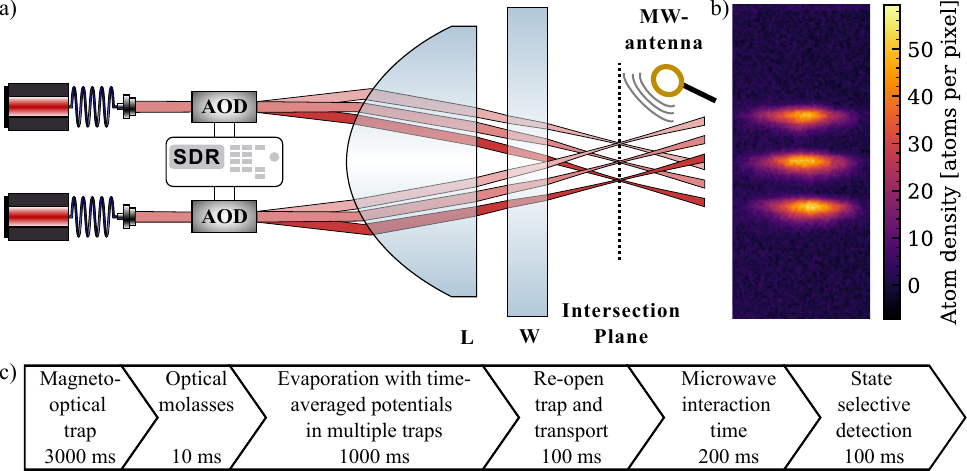}
    \caption{(a) Sketch of the key elements of the necessary setup. Two laser beams are each modulated by a two-dimensional acoustic optical deflector (AOD) driven by a software-defined radio (SDR) and focused on cold $^{87}$Rb atoms forming a crossed optical dipole trap (cODT). A microwave antenna provides a source to interrogate the atoms. (b) Fluorescence imaging of the resulting vertically separated traps after a ballistic expansion of 0.1\,ms. (c)  Experimental sequence with typical time scales used to obtain multiple cold atomic ensembles and perform clock-like measurements. There is an additional time of 4.7\,s between cycles leading to a total cycle time of 9.2\,s.} 
    \label{fig:setup}
\end{figure*}
\section{Setup}

The main requirement for this experiment is a setup capable of producing spatially separated cold atom ensembles trapped at different intensities by the same light source. The setup chosen for this demonstration was designed and built for the INTENTAS project and is described in large detail in Ref.~\cite{Anton24}, with additional description of the crossed-beam optical dipole trap (cODT) setup given in Ref.~\cite{Haase2025}. 
A combination of a two-dimensional magneto-optical trap (2D$^{+}$-MOT) with a 3D-MOT provides a source of up to 10$^{9}$ cooled $^{87}$Rb atoms. The atoms are subsequently cooled down to $\sim$40\,\textmu K using an optical molasses phase. A 1064\,nm crossed-beam optical dipole trap (cODT) is then used to trap atoms and cool them down further via evaporative cooling. It is formed by sending each laser through a two-dimensional acousto-optic deflector (AOD) driven by a software-defined radio (SDR). Both beams are then focused into the vacuum chamber by the same large-aperture lens, overlapping at an angle of 30$^\circ$ at the location of the cooled atoms. Initially, the horizontal degree of freedom of the AODs is modulated in order to create time-averaged potentials, thus increasing effective trap volume and atom number. By applying multiple RF frequencies to the vertical degree of freedom of the AODs, separated traps along the vertical axis can be realized. After the initial loading of the individual traps, the modulation is reduced to compress the clouds, and the intensity is lowered to initiate evaporative cooling. Subsequently, the traps are again expanded using time-averaged box-shaped potentials (Fig. \ref{fig:box1}, \ref{fig:box2} and \cite{Trypogeorgos13}) in order to mitigate density-related frequency shifts and decoherence. By these means, multiple atomic clouds can be prepared, each containing about 10,000 atoms at about 1\,\textmu K. While the setup is in principle capable of Bose-Einstein condensation, this is not required for the present study. 
The system is equipped with a microwave antenna surrounding one of the viewports, which offers the possibility of coherent excitation of the hyperfine clock transition of $^{87}$Rb. The antenna is driven by a low-phase-noise microwave source built according to Ref.~\cite{Meyer-Hoppe23}. In order to ensure long-term stability, the source is referenced via GNSS. 
For a state-selective detection perpendicular to the separation axis, one of the coils of the 3D-MOT can be used to apply a magnetic gradient field of $100$~mT/m.
The gradient separates the atoms according to their Zeeman state during ballistic expansion prior to detection. 
The key elements of the experimental setup, the basic experimental sequence in order to create, interrogate and detect the atoms, and an example image of the atoms are illustrated in Figure \ref{fig:setup}.

\section{Experimental results}

\begin{figure}[htbp]

    \centering
    \includegraphics[width=1\linewidth]{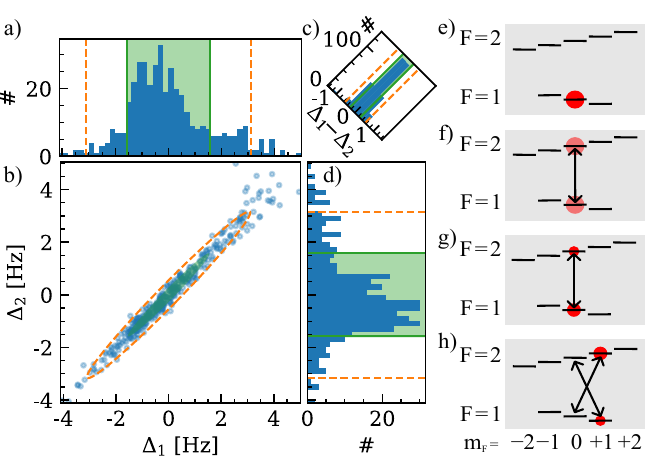}
    \caption{(b) Measurement of the detuning of trap-2 as a function of the detuning of trap-1. The green (red) ellipse indicates the $1(2)\,\sigma$ contour of a double Gaussian with a correlation angle. (a) + (d) Projection on the respective trap axis showing measurements per detuning as a histogram. (c) Fluctuations of the difference between the two detunings. (e)-(h) Total sequence of preparation and detection.} 
    \label{fig:seq}
\end{figure}

In order to demonstrate the concept, we choose a vertical separation of 170\,\textmu m between individual atomic ensembles trapped in horizontally expanded traps with a width of about 550\,\textmu m. The atoms in these traps are then prepared purely in the clock ground state $|F, m_F \rangle = |1,0\rangle$ by removing the atoms from $|1,\pm1\rangle$ using microwave transitions to the F=2 hyperfine manifold and subsequent resonant light pushes. The microwave is then used to drive the clock transition ($|1,0\rangle$ to $|2,0\rangle$) in a Ramsey sequence. The transition frequency is given by $f_0$(I)\,=\,$f_0$\,+\,$\alpha$I, with $f_0$, the atomic unperturbed transition frequency, and $\alpha$, the DLS coefficient encapsulating the intensity($\text{I}$)-dependent shift. 
In order to detect the two output states simultaneously, the atoms are transferred using Landau-Zener transitions from the magnetic-insensitive states into the magnetic-sensitive state of the other hyperfine manifold ($|2,0\rangle$ to $|1,+1\rangle$ and $|1,0\rangle$ to $|2,+1\rangle$). The sequence is shown in Fig.~\ref{fig:seq}. By applying a magnetic field gradient to split the Zeeman states, the two atomic ensembles per trap can be distinguished on the image. Due to technical limitations for switching times set by the coil, the magnetic field gradient was already applied during the final trapping phase. The atoms are released subsequently with the magnetic gradient field still applied. This enables fluorescence imaging with sufficient separation of the atomic clouds after 1.1\,ms of ballistic expansion. 
A Ramsey sequence, two $\frac{\pi}{2}$-pulses separated by a Ramsey time $T$, is chosen to perform transition-frequency measurements on the clock transition. To perform a spectroscopy, a variable microwave frequency is applied with a fixed Ramsey time and the population transfer to the excited state is measured. This provides the resonant frequency $f_0$(I) depending on the intensity $I$ of the trapping light in the trapped-atom configuration due to the detuning caused by the DLS. 
To perform frequency-shift-sensitive measurements in every shot, the microwave frequency $f_\text{mw}$ is detuned from the resonance by $\pm \frac{2n+1}{4T}$ with $n \in \mathbb{N}_0$, typically $n = 0$. At these so-called mid-fringe points, the two states are equally populated and the frequency sensitivity is maximized. The measured population imbalance is therefore used to determine the detuning $\Delta$ between the applied microwave frequency $f_\text{mw}$ and the determined resonance frequency $f_0$(I). 
In a first measurement series, we probe the amount of common-mode and local fluctuations.
To this end, two traps are operated in a symmetric configuration and interrogated with the same mw field and a Ramsey time of $T=25$\,ms.
Figure~\ref{fig:seq} shows the recorded population transfer within the two traps (referred to as trap-1 and trap-2 counting from the top) for $412$ consecutive measurements. 
Fluctuations, as given by the width of the distribution for each individual trap, amount to $\sigma(\Delta_{1}) = 1.55 \pm 0.05$\,Hz and $\sigma(\Delta_{2}) = 1.58 \pm 0.06$\,Hz.

Obtaining the difference between the two values yields a left-over fluctuation of $\sigma(\Delta_{1}-\Delta_{2}) = 0.275\pm 0.010$\,Hz.
This is illustrated by projecting the distribution to the major axes of the ellipse in Figure~\ref{fig:seq}.

The main common-noise source are fluctuations of the magnetic field, as the experiment did not have any passive shielding or active B-field stabilization during the presented measurements.
This is backed by an independent measurement of the magnetic‑field fluctuations on magnetically sensitive atomic transitions, which yield $\Delta B = 0.123 \pm 0.003$\,\textmu T, corresponding to frequency fluctuations of $\Delta f = 1.39\pm 0.03$\,Hz caused by the quadratic Zeeman shift and thus dominating the common‑mode noise. As this measurement was taken in a separate campaign, the quoted uncertainty is statistical only and does not cover day-to-day variations of the ambient field. Accordingly, the observed common-mode noise amplitude would require fluctuations of $\sim$\,10\% above the independently determined value.

The fluctuations of the frequency difference have multiple contributions.
Firstly, fluctuating atom numbers lead to changing density shifts.
Experimentally, we obtain an atom number dependent shift of -0.27$\,\pm\,$0.03\,mHz per atom, which we attribute to the density shift.
Typically, the atom number difference in the symmetric trap configuration has been measured to be around 400 atoms, which corresponds to fluctuations of 0.1\,Hz.
Secondly, the shot noise fluctuations translate to an uncertainty of the frequency measurement of 0.09\,Hz. However, our detection noise in this case is not able to operate at the standard quantum limit and technical noise is still present in the setup. We estimate the influence of the non-common detection noise in an independent measurement with $\pi/2$ beam splitter pulses and obtain a contribution of 0.3--0.4\,Hz.
The value is slightly too large due to additional technical noise in the beam splitter measurement, but indicates that detection noise is the major noise source for relative measurements in our system. 
 An analysis of the stability over time shows that the noise sources seem to average down over time in the observed window, see Figure \ref{fig_appendix_allan_deviation} in the Appendix for details.

\begin{figure}[htbp]

    \centering
    \includegraphics[width=0.49\linewidth]{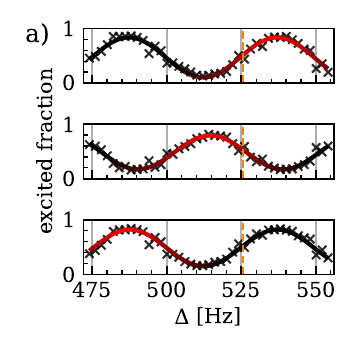}
    \includegraphics[width=0.49\linewidth]{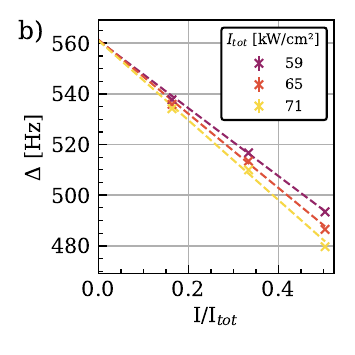}
    \caption{(a) The fraction of atoms in the excited state as a function of frequency detuning with a Ramsey time T = 20\,ms. The red marked region depicts the zeroth fringe and the orange dashed line shows where the microwave frequency is set to for simultaneous measurements of all three traps. The overall intensity used here is 65\,kW/cm$^2$ (b) Frequency detunings calculated from Ramsey scans as shown in panel (a)  vs intensity ratio for three different set total intensities (59, 65 and 71\,kW/cm$^2$). The red data points are taken from the fit to the measurements shown in (a).}
    \label{fig:results_ramseyscan}
\end{figure}

Since magnetic‑field fluctuations dominate the common-mode noise in our system, which can be shielded without significant efforts, we deliberately vary the total dipole‑trap power to demonstrate our suppression method.
Hereby, the AODs are used to create multiple optical traps with different laser intensities.
Each AOD is driven by an rf signal containing several frequencies; the relative power of these frequencies can be set precisely and remains stable over time.
By varying the total optical power, we emulate the power fluctuations that our technique is intended to suppress.
In our realization, we chose three traps (trap-1, trap-2 and trap-3) with intensities $I_{1,2,3}$ set by fixed ratios of the total intensity, $I_n = \frac{n}{6} I_{\text{tot}}$. 
In this configuration, fluctuations of the total trap intensity contribute as common-mode shifts and can be removed by extrapolation, such that the residual first-order uncertainty is determined by fluctuations of the relative intensity ratios set by the rf chain.
As we operate in a box-shaped potential, this should in first order only affect the DLS and not collisional shifts nor other trap-geometry-related effects.
The measured residual differential fluctuations of the individual traps in the measurement presented in Fig. \ref{fig:seq} of $\sigma_n = \sigma(\Delta_1-\Delta_2) / \sqrt{2} = 0.194$\,Hz scale due to the extrapolation in the used configuration with $\sqrt{7/3}$ assuming independent noise sources. Additionally a scaling by a factor of $25 \text{ms} / 20 \text{ms}$  due to an adapted interrogation time can be expected. We therefore expect a contribution of $0.37$\,Hz on $\sigma_0$. This assumes that the per-trap differential noise, dominated by detection, is independent of trap intensity and thus transfers from the symmetric to the asymmetric configuration.
First, we perform a full frequency scan for a given overall power (exemplary results in Figure~\ref{fig:results_ramseyscan}a) to obtain the resonance frequency for each trap. As expected, the resonance frequencies show a linear, intensity-dependent shift.
From the linear extrapolation, we obtain DLS-free resonance frequencies (Figure~\ref{fig:results_ramseyscan}b).
By repeating the scans with different total intensities I$_\text{tot}$, it is validated that the extrapolation yields the same value within uncertainties as shown in Table~\ref{tab:f0}. The uncertainty is given by the linear fitting to the three asymmetric traps, where each data point's uncertainty is estimated by the fit uncertainty of the underlying Ramsey scan.
\begin{table}[h]
\centering
\begin{tabular}{c|c}
\textbf{$\text{I}_\text{tot}$ [kW/cm$^2$]} & \textbf{$f_0(\text{I}=0)$}  [Hz] \\ \hline
         59      &  6\,834\,683\,172.0\,$\pm$\,0.6                 \\
         65      &  6\,834\,683\,172.3\,$\pm$\,0.4                  \\
         71      &  6\,834\,683\,172.0\,$\pm$\,0.5                   \\
\label{tab:f0}
\end{tabular}
\caption{Extrapolated values of the DLS-free frequency $f_0(\text{I}=0)$ for different overall intensities I$_\text{tot}$ taken from Ramsey scans shown in Figure \ref{fig:results_ramseyscan}. These intensities have been calculated from approximated expected intensities in the box-shaped trap potential at the location of the atoms given the setting for the overall power in the dipole trap as defined by the calibrated intensity stabilization.}
\label{tab:f0}
\end{table}

Comparing the obtained result to the recommended value for the transition 6\,834\,682\,610.9\,Hz \cite{margolis2024cipm}, we find a mean deviation of 561.2\,$\pm$\,0.5 Hz. The collisional shift for 10$^{4}$ atoms is expected to contribute additionally by -2.7\,$\pm$\,0.3\,Hz. The residual unaccounted shift of 563.9\,$\pm$\,0.6 Hz would correspond to a magnetic field of 99.02\,$\pm$\,0.05\,\textmu T. This is close to the target value of 100\,\textmu T set by the coils given by a rather large uncertainty on the calibration factor at the time of the measurements. Notably, in a future iteration this value could be determined using either an external sensor or utilizing the magnetic-sensitive transitions within interleaved or possibly simultaneous measurements.
While these results demonstrate that the DLS can be determined through extrapolation of intensity-varied traps, the presented analysis relies on data accumulation and retrospective fitting procedures.

\begin{figure*}[htbp]

    \centering

    \includegraphics[width=0.8\textwidth]{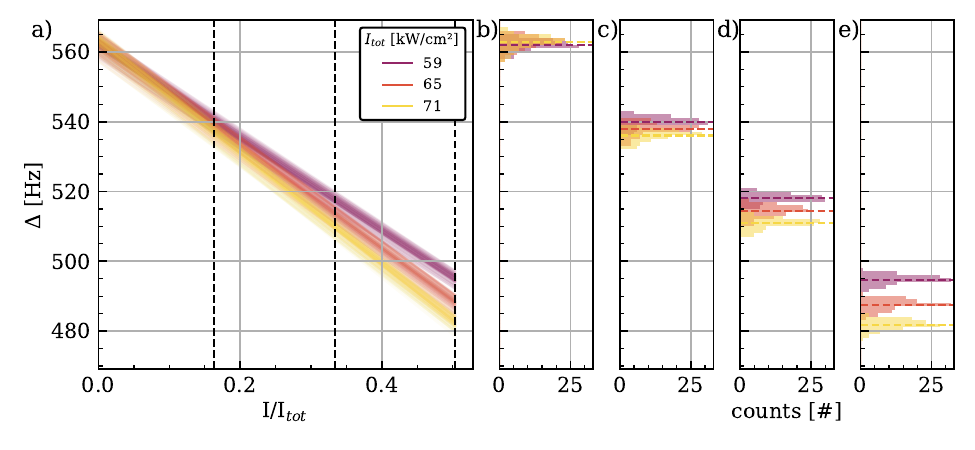}
    \caption{(a) Multiple individual runs with three traps at fixed intensity ratios (dashed black line) but varied overall intensity (colors) using a Ramsey time of 20\,ms and the calibration from the data underlying Figure \ref{fig:results_ramseyscan} (b) Histogram for the extrapolated values at $f(I=0)$ for all three overall intensities. (c-e) Histogram for the measured detunings at the respective intensity for a given ratio of the overall intensity I$_\text{tot}$} 
    \label{fig:results3}
\end{figure*}

Now, we demonstrate that the method can also be employed to obtain DLS-free frequency estimators for each individual experimental run.
In our case, this requires matching Ramsey time and intensity ratios such that the frequency shift between the individual traps is a multiple of $\frac{1}{2\cdot T}$. The frequency differences between the traps are determined beforehand by Ramsey scans with stepwise increasing T, to ensure an unambiguous assignment of the central fringe. In such a configuration, the mw frequency can be chosen to enable simultaneous mid-fringe measurements for all traps. Here, the intensity ratios are fixed by the rf amplitudes set at the SDR, so the condition is met by adjusting the Ramsey time to 20\,ms (Figure~\ref{fig:results_ramseyscan}).
Figure~\ref{fig:results3} shows the result of a measurement run, where a DLS-free frequency is extracted from each individual realization.
Artificial intensity fluctuations of $\pm9\,\%$ around the central setting are implemented by repeating the measurements at three overall intensities with the same fixed intensity ratios for the three traps. This exceeds the residual power fluctuations of a stabilized dipole trap by more than two orders of magnitude, and shifts the highest-intensity trap by $\pm6.5$\,Hz, several times the single-shot width of the extrapolated frequency. The fluctuations of the extrapolated value $f_0(\text{I}=0)$ amount to 1.74\,$\pm$\,0.14\,Hz, with individual values of 1.46\,Hz, 1.85\,Hz and 1.92\,Hz for the three overall intensities. This is in agreement with 1.59\,Hz given by the common-mode noise and the expected 0.37\,Hz differential contribution.

For $n=100$ shots per setting, the extrapolated DLS-free estimates for the mean frequency are given in Tab. \ref{tab:f1}. While the mean obtained at the intermediate and high overall intensities agree within statistical uncertainty, the lowest overall intensity setting shows an offset of about $0.7\text\,\mathrm{Hz}$. While its statistical significance is about 3$\sigma$, taking into account the systematic uncertainty from the underlying calibration of population imbalance to frequency shift for each overall intensity setting, reduces the significance such that it agrees within uncertainties. However, if there is an additional systematic effect at reduced trap depth, such as deviations from strictly proportional intensity scaling or changes in ensemble properties affecting residual density-dependent shifts, a better calibration would be necessary to resolve the influence. From the data given the accuracy of the method is therefore limited to the 10$^{-10}$ level. Notably, the data in Table \ref{tab:f0} is shifted by about 1.37\,Hz from the mean result in Table \ref{tab:f1}. We attribute this to a drift of the magnetic field on the order of 0.1\,\textmu T.
Future work investigating the limits of the cancellation performance will require to study this systematically, albeit in more realistic scenarios with smaller intensity fluctuations.

\begin{table}[h]
\centering
\begin{tabular}{c|c}
\textbf{$\text{I}_\text{tot}$ [kW/cm$^2$]} & \textbf{$f_0(\text{I}=0)$}  [Hz] \\ \hline
         59      &  6\,834\,683\,173.0\,$\pm \, 0.15\,(\text{stat}) \pm \, 0.6\,(\text{syst})$                \\
         65      &  6\,834\,683\,173.7\,$\pm \, 0.19\,(\text{stat}) \pm \, 0.4\,(\text{syst})$                 \\
         71      &  6\,834\,683\,173.7\,$\pm \, 0.19\,(\text{stat}) \pm \, 0.5\,(\text{syst})$                 \\
\end{tabular}
\caption{Extrapolated values of the DLS-free frequency $f_0(\text{I}=0)$ for different overall intensities I$_\text{tot}$ taken from single shot extrapolations shown in Figure \ref{fig:results3}. The systematic uncertainties are given by the individual calibration of the Ramsey scans for each particular overall intensity setting.}
\label{tab:f1}
\end{table}

\section{Outlook}

\begin{figure}[H]

    \centering
    \includegraphics[width=0.9\linewidth]{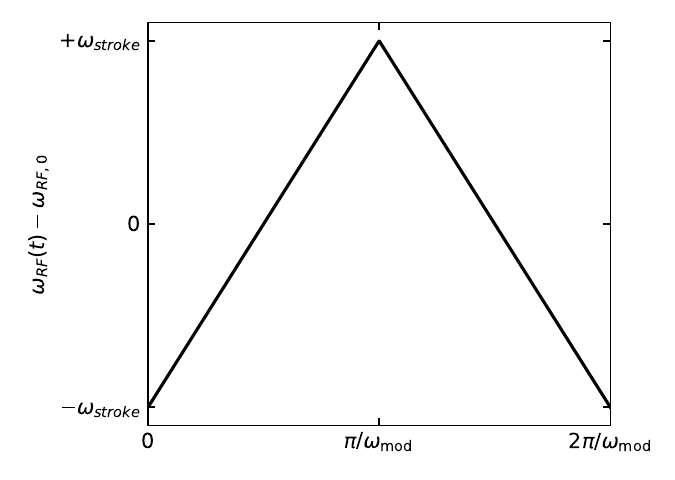}
    \caption{Frequency modulation for the generation of box potentials: Comparable to \cite{Roy16} the generation of box potentials as described in \cite{Trypogeorgos13} can be realized by frequency modulation of the RF signal sent to the AODs. During one period of the modulation $2 \pi / \omega_{mod}$, the frequency of the driving rf signal is linearly swept from $-\omega_{stroke}$ to $+\omega_{stroke}$ and back.}
    \label{fig:box1}
\end{figure}

\begin{figure}[H]
    \centering
    \includegraphics[width=0.9\linewidth]{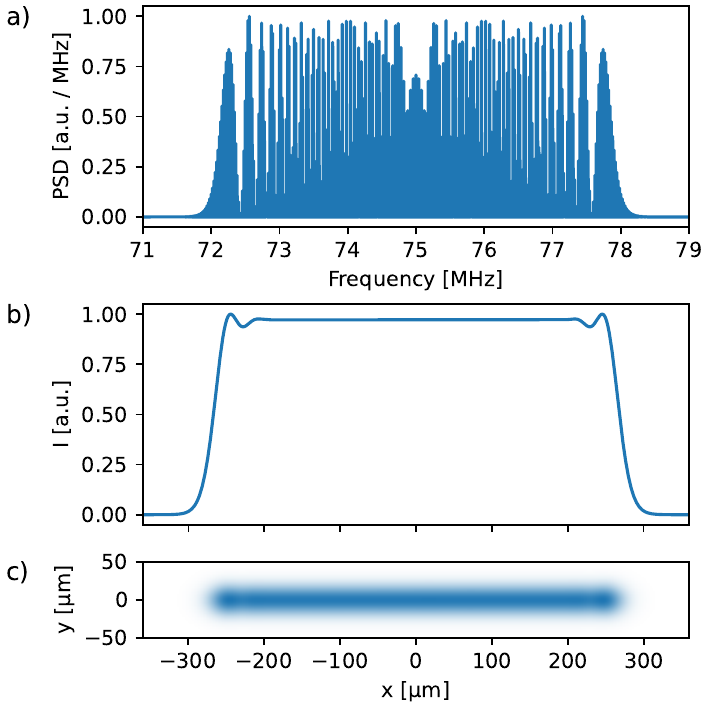}
    \caption{Simulated power spectral density (PSD) for the used box potential with $\omega_{stroke} = 3$\,MHz of the rf signal (a) and the laser light intensity in 1D (b) and 2D (c). In panel (c) white areas correspond to low intensity while blue areas correspond to high intensity.}
    \label{fig:box2}
\end{figure}

\begin{figure*}[htbp]

    \centering

    \includegraphics[width=0.8\textwidth]{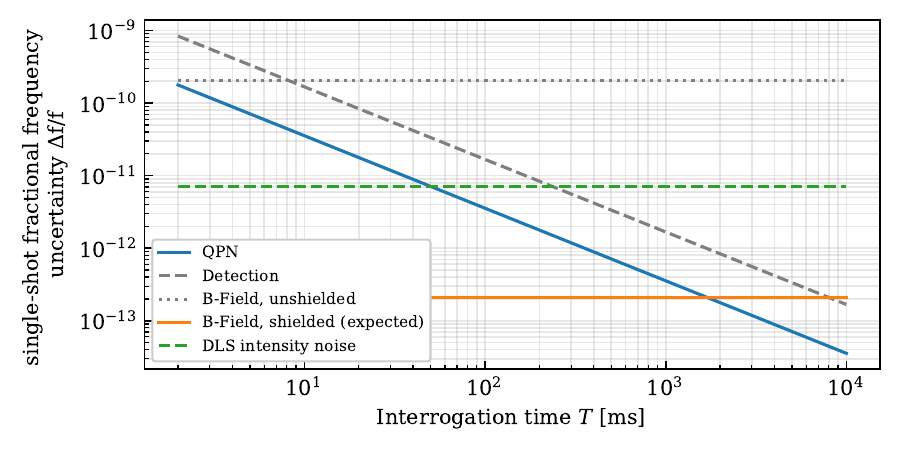}
    \caption{Estimated single-shot fractional frequency uncertainty
  $\Delta f/f$ of a single measurement as a function of the
  Ramsey interrogation time $T$. The quantum projection noise (QPN) is
  given for an extrapolation from three traps with $10^{4}$ atoms each.
  The detection noise corresponds to the 0.37\, Hz expected for the asymmetric configuration at 20 ms interrogation time and scales with 1/T. Magnetic-field noise is shown for the present,
  unshielded operation ($2\times10^{-10}$) and for the expected
  performance with the planned shielding (measured shielding factor
  $>1000$). The DLS contribution assumes a relative intensity error of
  $10^{-3}$ at the location of the atoms; it scales
  linearly, giving $7\times10^{-13}$ at $10^{-4}$. This error is not
  identical to the performance of an intensity stabilization, since beam
  pointing and the ensemble temperature also change the intensity
  experienced by the atoms. Contributions that are independent of $T$
  appear as horizontal lines; where they cross the QPN, a longer
  interrogation time no longer improves the single-shot resolution. The
  curves show single-shot uncertainties. Slow drifts of the trap intensity do not average down and if not accounted for
enter the accuracy budget. Note that the interrogation time that can
  actually be realized is separately bounded by the coherence time. Contributions specific to the
  present apparatus, such as the stability of the current driver for the
  quantization field and the phase noise of the microwave local
  oscillator entering through the Dick effect, are not included here.}
    \label{fig:app3}
\end{figure*}

The demonstrated cancellation of differential light shifts (DLS) using multiple, simultaneously interrogated atomic ensembles provides a versatile route toward improving the stability and accuracy of trapped-atom microwave clocks. 
In an optimized apparatus, the method removes the DLS contribution that is independent of the interrogation time and therefore sets a floor that no extension of the Ramsey time can overcome (Figure~\ref{fig:app3}). Exploiting this at long Ramsey times requires, in addition, a sufficient coherence time. The box-shaped potentials (Figure~\ref{fig:box1} and Figure~\ref{fig:box2}) applied here empirically have not been fully investigated in this respect, and a systematic study of the attainable coherence time in such potentials would be of interest.

For the presented setup additional mitigation of technical noise such as stray light and camera readout noise remains necessary, while the proof-of-principle realization showed that the residual fluctuations are dominated by ambient magnetic-field noise. With passive shielding or active stabilization and a reduction of the absolute field strength, these fluctuations can be reduced to the \textmu Hz level. The stability of the extracted frequency is then set by the remaining common-mode contributions (i.e. microwave phase noise) which pass through the extrapolation unchanged without degrading the cancellation. Differential noise sources such as detection noise, atom-number fluctuations and atom shot noise instead enter the extrapolation amplified and determine the precision with which the DLS is canceled in each shot; unlike a bias, they average down with measurement time. 
At the present level of detection noise, the achievable degree of cancellation is not yet resolved. Testing it below the $10^{-10}$ level demonstrated here requires also improved calibration statistics of the Ramsey scans underlying each setting and measurements at additional overall intensities, which would also test the linearity of the extrapolation beyond the three settings used here.

Ultimately, we expect that errors of the intensity ratios will limit the method: a static ratio error biases the extrapolated frequency and sets an accuracy floor, whereas a slow drift limits the long-term stability. The method does not remove the sensitivity to intensity errors, it transfers it from the absolute intensity at the location of the atoms, which includes beam pointing and ensemble temperature and is hard to control or even monitor, to the much better controlled intensity ratios, which are set digitally and can be monitored out-of-loop. The gain is given by the ratio of these two stabilities. Deviations can arise from the AODs, whose diffraction efficiency is nonlinear in rf power and thermally sensitive, from the rf amplifiers, or from the rf synthesis itself, and require careful characterization. However, possible out-of-loop monitoring strategies of these parameters can then push the limits of the method further. 

The underlying concept is not restricted to DLS. Any systematic shift that can be tuned independently in different ensembles becomes accessible to in-situ extrapolation. Varying trap depth and atom number enables characterization and suppression of density-dependent collisional shifts. In extended geometries, such as two-dimensional arrays of time-averaged traps~\cite{Haase2025}, multi-parameter extrapolation becomes feasible, allowing simultaneous removal of several systematic contributions within a single experimental cycle.
The presented method could also provide technical insights into the optical dipole trap itself, when operated with a reversed logic it can be used to characterize the system regarding the intensity at the very location of the atoms in real-time. Conversely, if the absolute intensity at the atoms is independently calibrated, the same measurement yields the differential polarizability $\delta\alpha$, for which to our knowledge only a single measurement with 14\% uncertainty exists \cite{KleineBuening2011}. The present data are consistent with that value, but the absolute power calibration and the characterization of the beam overlap at the focus do not support a competitive uncertainty. An improved determination would be valuable for the calibration route discussed above, which the method presented here avoids by construction.

\section{Conclusion}

We have demonstrated a technique to cancel the influence of differential light shifts (DLS) in optical dipole traps by simultaneously interrogating multiple atomic ensembles at different trap intensities, set by fixed ratios from a common light source. Although the method was implemented here using $^{87}$Rb, it is general and does not rely on magic wavelengths, species-specific cancellation strategies, nor on knowledge of the absolute intensity at the atoms. Using time-averaged potentials, arrays of spatially separated traps can be created, enabling strong common-mode noise rejection while providing independent control of the local trap intensity. By operating these traps at different intensities, we obtain an extrapolation to a DLS-free frequency for each individual experimental run. Introducing artificial light fluctuations of $\pm9\,\%$ of the total trap intensity, confirmed that the DLS can be canceled in situ on a shot-to-shot basis, effectively removing the dominant coupling between trap intensity noise and the clock transition. 
The extrapolated DLS-free frequency is reproducible at the $10^{-10}$ level, at present limited by detection noise and by the calibration statistics of the underlying Ramsey scans rather than by the cancellation itself. The scheme thereby addresses the stability and accuracy limitations imposed by the trapping light, whereas the decoherence caused by its inhomogeneity has to be treated separately, here by shallow box-shaped potentials.
Beyond DLS, the approach can be generalized to other systematic shifts. By selecting specific densities in each trap, density-dependent frequency shifts and fluctuations can likewise be characterized and suppressed through real-time extrapolation. These results establish a flexible and powerful framework for improving the accuracy and stability of trapped-atom clocks and related quantum sensors.

\section*{Declarations}

\subsection*{Availability of data and materials}
Data are available upon reasonable request.

\subsection*{Competing interests}
The authors declare that they are inventors on a patent related to the device presented in this work.

\subsection*{Authors contributions}
A.F. drafted and edited the manuscript. J.S.H contributed to the editing, writing and figures.  C.K., J.K., I.B. contributed to the writing. J.S.H, C.K. conceived the original concept. J.S.H and A.F. performed the measurements and data analysis. All authors reviewed the manuscript.

\subsection*{Funding}
This work is supported by the German Space Agency (DLR) with funds provided by the BMWK
under Grant No. 50WM2174. Supporting work was also contributed by the Deutsche Forschungsgemeinschaft (DFG) under
Germany’s Excellence Strategy within the Cluster of Excellence QuantumFrontiers (EXC
2123, Project ID 390837967).

\subsection*{Acknowledgments}
We want to thank Janina Hamann for technical work on parts of the setup used for this study.

\clearpage

\subsection*{Appendix}

\begin{figure}[H]
    \centering
    \includegraphics[width=0.9\linewidth]{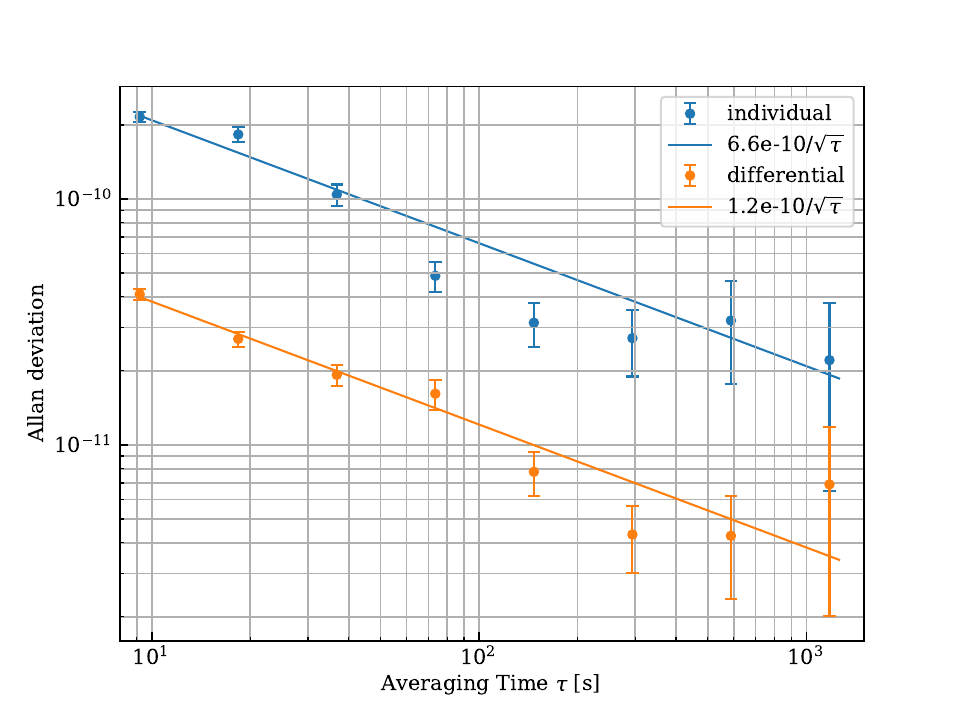}
    \caption{Allan deviation of the data shown in Figure \ref{fig:seq}. Lines show the expected $1/\sqrt{\tau}$ behavior. The data average down over the measurement period. The differential measurement has a lower initial Allan deviation as it suppresses common noise sources.}
    \label{fig_appendix_allan_deviation}
\end{figure}

%\begin{acknowledgments}
%Many Thanks
%\end{acknowledgments}
\clearpage
\bibliography{AC-Stark.bib}% Produces the bibliography via BibTeX.

\end{document}